# A Novel Capacitance-to-Phase Transducer Based on a Simple Resonant Circuit

**Frank J. van Kann[1] and Alexey V. Veryaskin[2]**

[1] Department of Physics, School of Physics, Mathematics and Computing, University of Western Australia, 35 Stirling Highway, Crawley, WA 6009, Australia

[2] Trinity Research Lab/QDM Lab, School of Physics, Mathematics and Computing, University of Western Australia, 35 Stirling Highway, Crawley, WA 6009, Australia

E-mail: alexey.veryaskin@uwa.edu.au

## Abstract

A novel room temperature capacitance-to-phase transducer is described, which uses a modified All-Pass filter architecture combined with a simple series resonant tank circuit with a moderate Q-factor. It is fashioned from a discrete inductor with small dissipation resonating with a grounded capacitor acting as the sensor, to obtain a resolution of $\Delta C/C \leq 10^{-11}$ in a 1 Hz bandwidth. The transducer converts the change in capacitance to the change in the phase of a carrier signal at the resonance frequency and is configured to act as a close approximation of the almost ideal All-Pass filter. This cancels out the effects of amplitude modulation when the carrier signal is imperfectly tuned to the resonance. When used as a mechanical displacement-to-phase transducer, the predicted position noise was approximately 3 femtometre/√Hz. Other applications of the proposed circuit include the measurement of the electric field, where the sensing capacitor depends on the applied electric field. In addition, the circuit can be easily adapted to function with very small capacitance values (1 - 2 pF) as is typical in MEMS-based transducers.

## 1. Introduction

The ability to measure minute variations of electrical capacitance has led to many academic and industrial applications including fundamental research and defence. Devices based on variable capacitors have been under development for many decades in areas such as precision accelerometry, gravimetry, gravity gradiometry, proximity sensing and many more [1, 2]. Precision MEMS accelerometers based on the moving plate capacitors technology have established their presence almost everywhere from smartphones and robotics to strategic defense applications and space missions ( as an example see [3] ). In a transducer, the capacitor becomes a variable capacitor when one of its electrodes is free to move under external disturbances whereas the other is fixed. This relative motion changes the value of the mutual capacitance and acts as a measure of external factors that cause disturbances. A capacitor can also change its value if electric charge is applied to one of its electrodes. Such capacitors are known as varactors [4]. By measuring this capacitance change it is possible to measure the applied electric field [5]. The change in capacitance must be translated into a measurable quantity – the amplitude, frequency or phase of a DC, audio, RF, microwave, or optical signal. DC-biased capacitive transducers are reciprocal devices [6], and while they have poor performance for low-frequency or DC signals, they are excellent for AC signals [7]. Piezo-electric accelerometers also belong to this category, where the DC bias field is provided by the crystal lattice, rather than an external voltage. In contrast, parametric transducers use AC drive signals and audio-frequency-based capacitive transducers have been known for approximately a century. In most cases they use capacitance bridge architecture, typically containing four capacitors at least one of which is a variable capacitor [8, 9]. RF bridges, using a differential transformer on resonance offer even greater performance [10 - 13]. Parametric capacitive sensors, which exploit the phase shift produced by modulating the capacitance in a resonant circuit, exhibit the greatest precision, when driven by an RF or microwave carrier signal at the resonance frequency [14, 15]. The minute changes in phase can be detected using phase-locked loops (PLLs) and/or using an interferometric read-out where one arm of an interferometer contains the variable capacitor (DUT) while the other contains the reference one. Interferometric measurements provide the most sensitive instrumentation for the detection of ultra-small phase differences in otherwise identical carrier signals propagating through arms of an interferometer [16 - 18]. Other practical phase measurement solutions aiming at precision capacitive sensing based on phase sensitive detectors have also been described [19] but not all of them allow for the use of grounded sensing capacitors which are required for most mechanical transducers. Matko and Milanović described a temperature-compensated capacitance-to-frequency converter with a claimed resolution of ±20 zF (1 zF = $10^{-21}$ Farad) [20]. However, this approach uses a floating variable capacitor and cannot be easily adapted for ultra-precision mechanical displacement measurements. A very high Q parallel resonant LC-tank (Q ≥1000) has been used as a simple circuit for either amplitude or phase sensitive detection of ultra-small capacitance variation [14, 15], but this is possible only at cryogenic temperatures to avoid being dominated by the Johnson noise at the resonance. Moreover, the carrier signal must be extremely stable and tuned exactly to the resonance to avoid significant amplitude modulation, which produces errors in the phase detection. The same considerations apply to high-Q active notch circuits using the resonant dip instead of resonant peak [21, 22]. In addition they significantly attenuate the carrier signal which is a problem if used in interferometers as the latter require large signals to provide optimum signal-to-noise ratio [17]. First order All-Pass filters based on a single operational amplifier (further down op amp) represent perfect capacitance-to-phase converters [23]. They have unity gain for all frequencies with no attenuation within the capabilities of the op amp being used. However, the capacitance-to-phase conversion rate is low in these first order filters in order to provide performance comparable to the best capacitive transducers mentioned above. Second order All-Pass filters based on a single op amp with a grounded capacitor, which can act as the sensing element, are not known. A detailed review of modern capacitive measurement techniques is presented in [24]. A novel approach to an ultra-sensitive room temperature capacitance-to-phase transducer using a grounded capacitor is presented below. At the heart of the proposed transducer is a second order All-Pass filter architecture based on a series LCR resonant notch such that the circuit retains the property of almost ideal All-Pass-like flat amplitude response while maintaining the capacitance-to-phase conversion rate of high-Q resonant tank circuits. The authors of this study found neither similar published proposals nor patent applications granted or expired worldwide. To initially simplify the transducer analysis, the circuit is coupled to an ideal amplifier, in the sense that it has infinite gain-bandwidth and input impedance, while retaining the usual noise sources associated with real amplifiers. The model was then extended to explore the effects of using a more realistic amplifier model.

## 2. The Simple Resonant Circuit as a Notch Constituent of the Capacitance-to-Phase Transducer

The essential resonant properties of the proposed transducer can be analysed using a circuit shown in Fig. 1. The circuit represents a Notch constituent of the complete transducer architecture described in Section 5. The series LCR resonant tank, consisting of inductor $L$ in series with the variable capacitor $C$, is coupled to the non-inverting input of an op amp. In the complete transducer architecture the capacitor C acts as the sensing element. The resistor $R_L$ is not a stand-alone component but represents the losses in the inductor, including winding resistance and magnetic losses. In this simplified model the inter-winding capacitance is also ignored but will be considered in Section 6. The amplifier is assumed to be ideal having infinite input impedance and infinite gain-bandwidth. However, the sources of noise associated with the amplifier are included in the model and consist of voltage noise $V_n$ and current noise $i_n$. These are stochastic signals and are characterised by the respective power spectral density (PSD) $S_{Vn}$ and $S_{in}$.

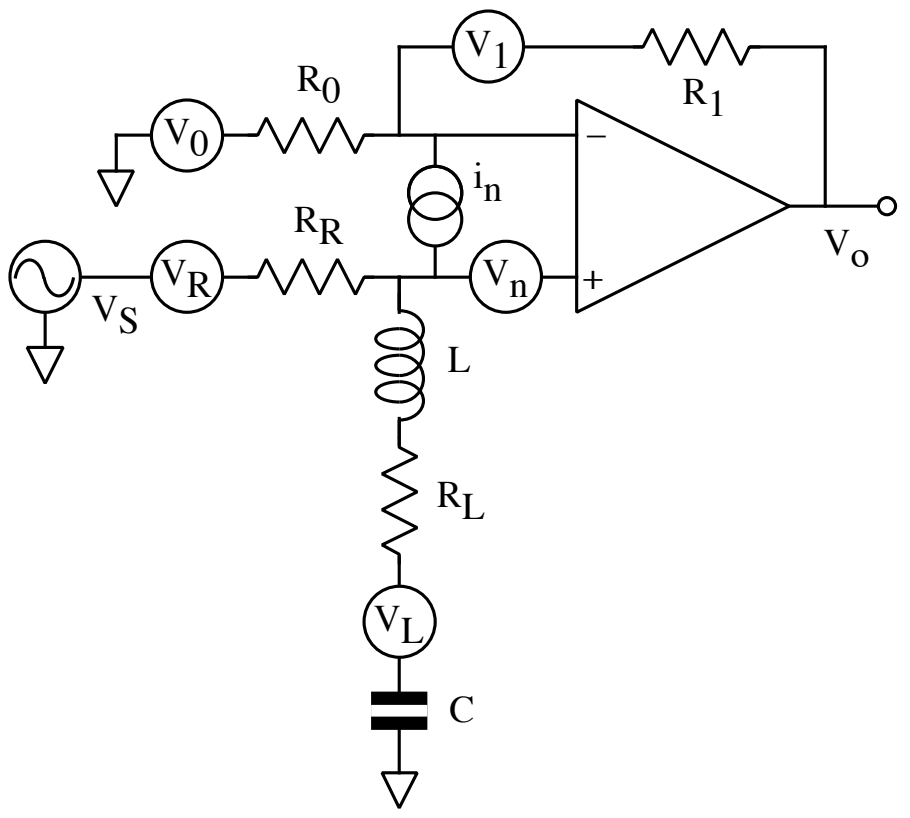

**Figure 1**. The schematic of the resonant Notch constituent of the proposed transducer showing the series LCR resonant tank coupled to the non-inverting input of the amplifier

Each resistor in the circuit is a source of Johnson noise, and the respective voltage sources shown in Fig. 1 are labelled with the same subscript as the corresponding resistor. Johnson noise sources are characterized by their respective PSD. The impedance of the circuit is illustrated in Fig. 2, which shows the complex impedance $Z_E(f)$ using the indicated component values, where $Z_E = s\,L + R_L + {}^1/_{s\,C}$ is the impedance of the LCR tank. Solid curves in Fig. 2 depict the impedance wherein the red and blue curves represent the magnitude and the phase respectively. The circuit is based around a high-Q 470 µH inductor, which is in series with the variable capacitor, whose equilibrium value is approximately 20 pF, to obtain a resonance within 1-2 megahertz range. The value of the resistor $R_L$ is not only the coil resistance, but corresponds to the observed Q. The PSD of the Johnson noise of the resonant circuit is proportional to the real part of the impedance $Re(Z(f))$, shown as the green curve in the figure and would be prohibitively large if operating at a resonance peak, using a different configuration of the LCR tank. An innovative solution is to configure the circuit to exhibit a resonance notch, where the slope of the phase as a function of frequency is equally large but the impedance is much smaller. The sensitivity of the transducer is proportional to this slope. The maximum slope occurs at the resonance frequency and the carrier signal $V_S$ (the "pump") should be set at that frequency.

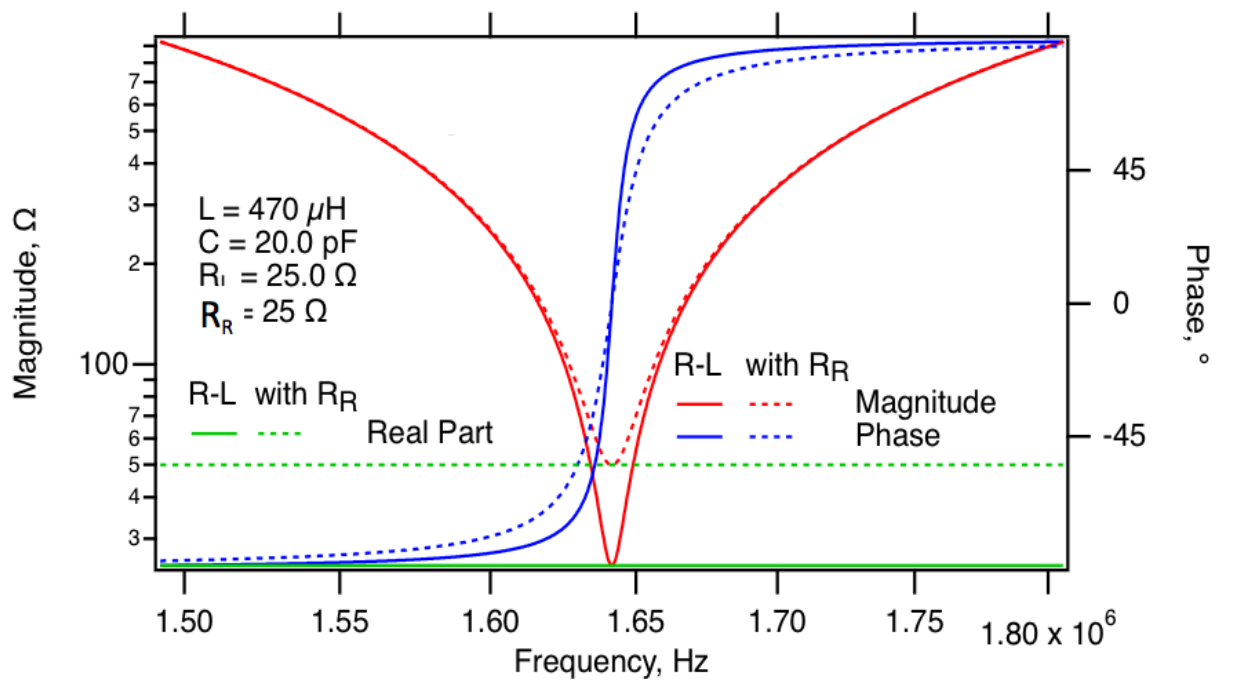

**Figure 2.** The impedance of the series LCR resonant tank, showing the magnitude (red), phase (blue) and real part (green), together with the impedance of the corresponding to the series resistor $R_R$ (dashed curves)

The behavior of the circuit is modified by the addition resistor $R_R$ in series, as shown in the circuit diagram of Fig. 1, and the corresponding impedance is also plotted in Fig. 2 as the dashed set of curves, labelled as "with $R_R$" and using the same color convention as before. The Q factor of the notch anti-resonance is visibly reduced, by exactly a factor of two in this example. The series combination of $R_R$ with the resonant circuit forms a voltage divider, the output of which is applied to the non-inverting input of the amplifier. The resulting transfer function of this combination is considered in the following section.

## 3. The Transfer Function of the Notch Constituent of the Capacitance-to-Phase Transducer

The transfer function between the signal source $V_S$ and the output of the amplifier is

$$H_R(f) = (1+g)\frac{Z_E}{Z_E+R_R} = (1+g)\frac{1+s\,R_L\,C+s^2L\,C}{1+s\,(R_L+R_R)\,C+s^2L\,C} \quad (1)$$

where above and further below $g = \frac{R_1}{R_0}$ is the absolute value of the inverting closed loop gain of the (ideal) amplifier. The definition of this transfer function assumes that other sources in the circuit (in particular the noise sources) are set to zero, and an analogous transfer function can be defined for each of these. The total signal is then obtained by linear superposition, or in the case of statistically independent stochastic signals, by summing the respective PSDs. The transfer function of equation (1) is that of an ideal notch filter, where the complex conjugate poles and zeros have the same frequency $f_o = 1/(2\pi\sqrt{LC_0})$, but different values of Q given for the zero and the pole respectively by

$$Q_Z = \frac{1}{R_L}\sqrt{\frac{L}{C}} \text{ and } Q_P = \frac{1}{R_L+R_R}\sqrt{\frac{L}{C}} \quad (2)$$

The transfer function at the output of the amplifier is plotted in Fig. 3, where the value of g is set to 3 (thereby setting the gain for the non-inverting input to g+1=4). The relative depth of the notch is $\frac{Q_P}{Q_Z}$ and the maximum slope of the phase occurs at the resonance frequency and is given by

$$f_o\frac{\partial\phi}{\partial f} = 2(Q_Z - Q_P) \quad (3)$$

This suggests making $Q_p$ as small as possible to maximize the slope and therefore the sensitivity. However, this also deepens the notch making the signal smaller and degrading the signal to noise ratio (SNR). That is the key to this circuit, i.e. operating at the notch would not be possible if the notch is too deep and there is a trade-off between a desired sharpness and an acceptable depth of the notch, as described in the next section.

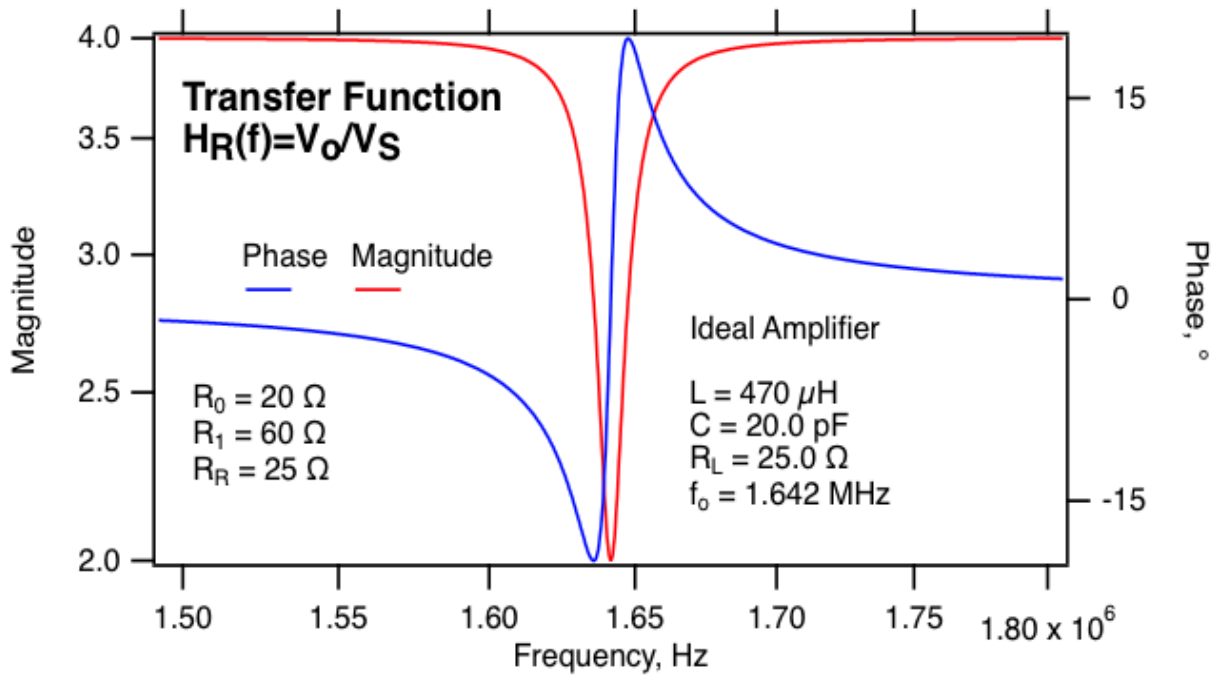

**Figure 3**. Transfer function of the Notch Constituent of the proposed Capacitance-to-Phase Transducer

## 4. Effective Noise of the Notch Constituent of the Capacitance-to-Phase Transducer

The noise of the circuit in Fig.1 is determined by both the phase noise of the pump signal and the noise of the amplifier. In practice, the phase noise of the pump signal dominates, but this can be greatly reduced using interferometric techniques as described in [17]. Assuming the latter is the case, the only noise limiting factor is the amplifier input noise. For small changes of capacitance around the average value, the PSD of the effective "capacitance noise" is

$$S_C = \left(\frac{\partial\varphi}{\partial C}\right)^{-2} S_\phi = \left(\frac{\partial f}{\partial C}\frac{\partial\varphi}{\partial f}\right)_{f_0}^{-2} S_\phi = \left(\frac{2C_o}{f_o}\frac{f_o}{2(Q_Z-Q_P)}\right)^2 S_\phi = \left(\frac{C_o}{Q_Z-Q_P}\right)^2 S_\phi \quad (4)$$

where $S_\phi$ the PSD of the phase noise of the circuit output signal and $C_o$ is the average value of $C$. Assuming that the pump signal is a pure, noise-free sinusoid, given by

$x(t) = A_o \, sin(2\pi f_o t)$

that passes through the transducer and the following amplifier and produces the desired phase with a fixed sensing capacitor $C_o$. To simplify the noise analysis, it is assumed that a noiseless narrowband phase detector, centered at the notch resonant frequency, is connected to the op amp's output. This means that the noise can be treated as a normal narrowband process described by Equation (5) :

$$y(t) = ASin(2\pi f_0 t + \varphi_0(C_0)) + n_\sigma(t) \quad (5)$$

where A is the cartier amplitude at the circuit output with a noiseless op amp, $\varphi_0$ is the phase shift and $n_\sigma(t)$ represents the output op amp noise with a standard deviation $\sigma_n$. The probability distribution of the phase of a cumulative signal similar to that of in Equation (5) was presented for the first time in [25] and is used in a different form below

$$P(\varphi - \varphi_o) = \\ = \frac{1}{\pi}\left\{ e^{-\frac{A^2}{2\sigma_n^2}} + \sqrt{\frac{\pi}{2}}\frac{A}{\sigma_n}\cos(\varphi - \varphi_o) e^{-\frac{A^2}{2\sigma_n^2}Sin^2(\varphi-\varphi_o)} \left[1 + \operatorname{erf}\left(\frac{A}{2\sigma_n}\cos(\varphi - \varphi_o)\right)\right]\right\}$$

where *erf(x)* is the error function.

The corresponding phase variance is as follows

$$\sigma_\varphi^2 = 2\int_0^\pi (\varphi - \varphi_0)^2 \, P(\varphi - \varphi_0) d(\varphi - \varphi_0)$$

The integral above can be accurately evaluated provided the following (close to real case) conditions are applied

$$\frac{A}{\sigma_n} \to \infty, \qquad \varphi - \varphi_0 \to 0$$

This yields the expected result :

$$\sigma_\varphi^2 = \frac{\sigma_n^2}{A^2}$$

then one has for the corresponding PSDs

$$S_\phi = \frac{S_n}{A^2}$$

It is interesting to note that there is a different approach to calculating the phase variance of the signal *y(t)* in Equation (5) based on the FFT applied to the latter as done in professional phase noise analysers [26]. The result turns out to be exactly the same as above.

Further one finds from Equation (4)

$$S_C = \left(\frac{C_o}{Q_Z - Q_P}\right)^2 \frac{S_{in}}{A_{in}^2}$$

where $S_{in}$ is the noise PSD and $A_{in}$ is the amplitude of the pump signal, both referred to the positive input of the amplifier. The notch attenuates the input signal, so that the amplitude at the input is $A_{in} = A\,\frac{Q_P}{Q_Z}$ and therefore

$$S_C = \left(\frac{C_o}{Q_Z - Q_P}\right)^2 \left(\frac{Q_Z}{Q_P}\right)^2 \frac{S_{in}}{A^2} \quad (6)$$

If the capacitor is used as a position transducer, the PSD of the effective displacement noise $S_x$ is given by

$$S_x = \left(\frac{\partial C}{\partial x}\right)^{-2} S_C = \left(\frac{x_o}{Q_Z - Q_P}\right)^2 \left(\frac{Q_Z}{Q_P}\right)^2 \frac{S_{in}}{A^2} \quad (7)$$

where $x_o$ is the average spacing between the capacitor plates when the motion is in the transverse direction. Therefore, while making $Q_P$ small does indeed increase the sensitivity, it also increases the noise and there is an optimum value, given by $Q_P = Q_Z/2$. This makes the optimum relative depth of the notch 50% or ~6 db. The value of $Q_Z$ should therefore be made as large as possible to minimize the noise, but this is limited by the properties of the inductor. It should be possible to achieve $Q_Z = 200$ at $f_o = 1MHz$. With such large Q-factors, the notch is quite sharp and there is a risk of amplitude modulation (AM) if the signal source is not perfectly tuned to the resonance. However, it is possible to transform the resonant circuit shown in Fig.1 into a second order All-Pass filter (APF) in a similar manner that a band-pass filter (BPF) can be transformed into an All-Pass Filter (APF) and this is described in the next section.

## 5. The Capacitance-to-Phase Transducer as a Second Order All-Pass Filter

It is well known that a second order BPF can be transformed into a second order APF using the formula $H_{\mathrm{APF}}(f) = 1 - 2\,\mathrm{H_{BPF}}(f)$ [27]. Similarly, the resonant Notch circuit in Fig.1 with the transfer function $\mathrm{H}_R(\mathrm{f})$ can be modified to produce the transfer function $H(f)$ defined by

$$H(f) = H_R(f) - g \quad (8)$$

Practically this weighted subtraction is accomplished by connecting the signal source to both $R_0$ and $R_R$ as shown in Fig. 4. This transformation leaves the frequency of both the complex conjugate poles and zeros and the Q of the poles unaltered but changes the Q of the zeros to $Q'_Z = \frac{Q_Z Q_P}{(1+g)Q_P - gQ_Z}$. The ideal APF corresponds to $Q'_Z = -Q_P$, which requires that the value of $g$ is set to $\frac{Q_Z+Q_P}{Q_Z-Q_P}$. When the notch filter is optimally tuned so that $Q_Z = 2Q_P$, then this requires that $g = 3$. This results in a perfectly flat transfer function, although in practice the tuning cannot be perfect but can nevertheless be made arbitrarily accurate by painstaking adjustment of the gain resistors. An example, where the error in the gain is ~0.1% is shown in Fig. 5. This reduces the AM sensitivity by three orders of magnitude.

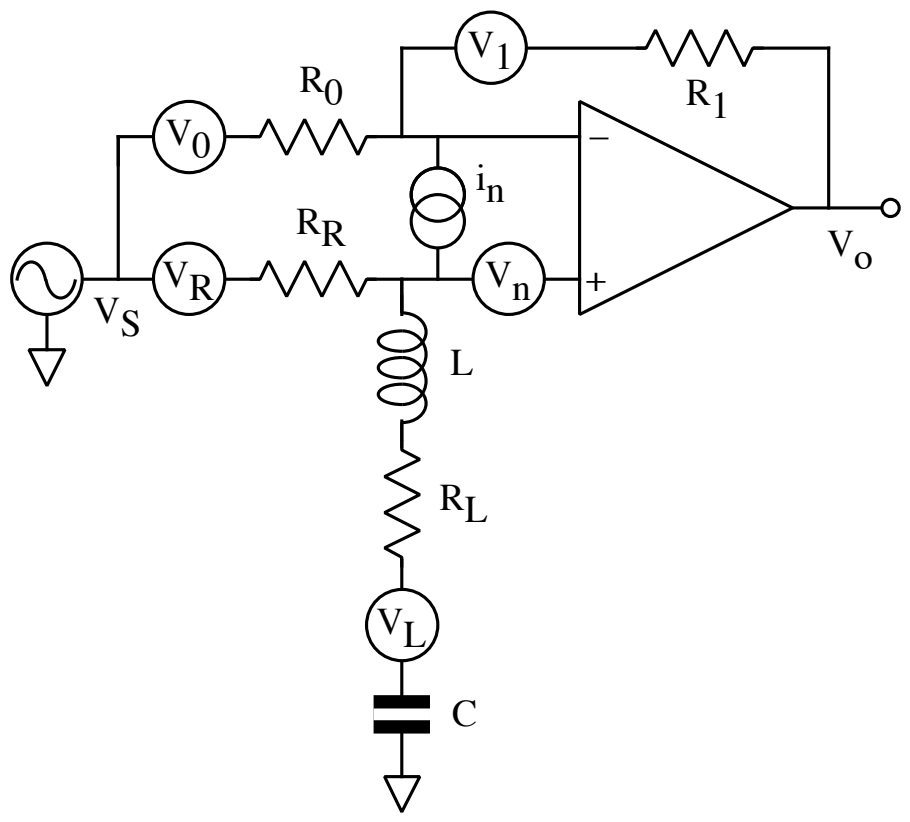

**Figure 4**. The complete circuit diagram of the proposed capacitance-to-phase transducer architecture

It is evident from the graph that the magnitude of the slope of the phase increases as the phase change across the resonance is increased to 360° and the slope is doubled from the original $Q_Z' = \frac{Q_Z}{2}$ of $H_R$ to $2Q_Z' = Q_Z$, resulting in a doubling of the sensitivity. However, the voltage noise of the amplifier is also doubled relative to the signal. The noise is increased by a factor $1 + g = 4$, whereas the signal is increased by a factor $g - 1 = 2$ and therefore the PSD for the capacitance measurement remains unchanged.

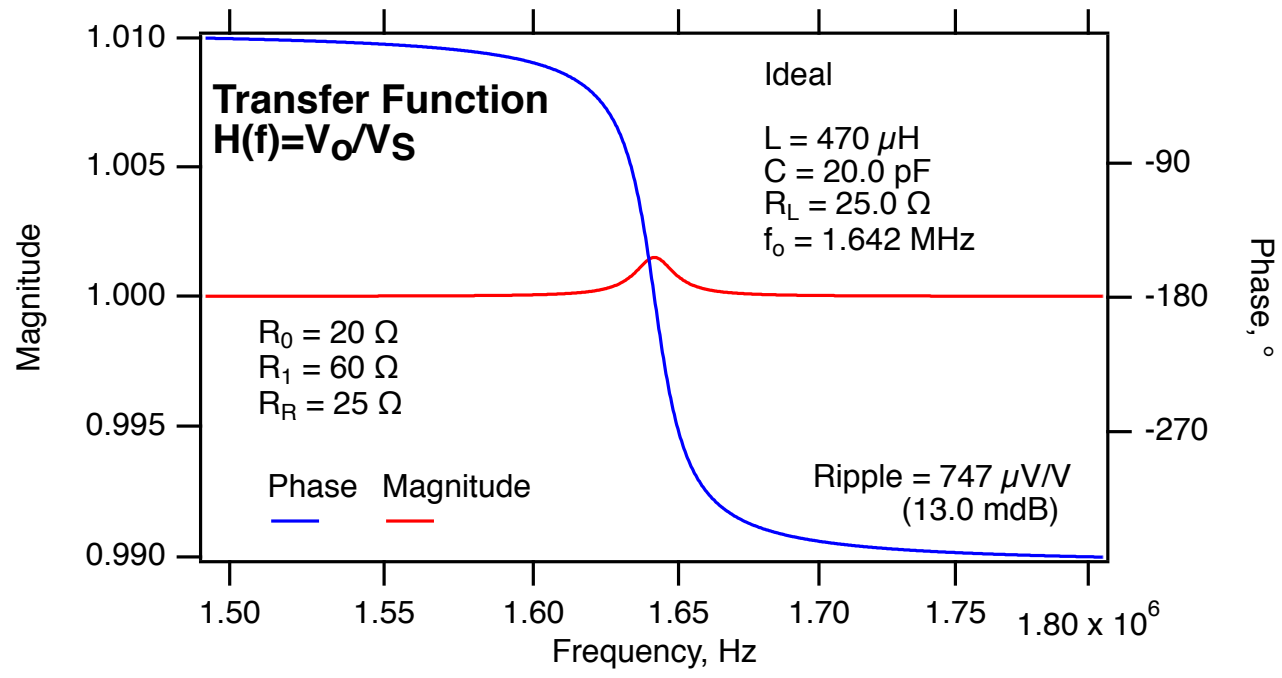

**Figure 5**. The transducer transfer function with tuning accurate to 0.1%

The slope of the phase curve can be expressed in terms of an effective Q-factor, using Equation (3), with $Q_{eff} = Q_Z - Q_P$ as shown in Fig. 6. Although the equation is valid only on resonance, it provides a measure of the sensitivity when the signal is slightly detuned from resonance. In practice, the departure from the ideal circuit for both the inductor and amplifier must be considered, to determine the impact on the performance which is considered in the following sections.

## 6. Extended Model of the Capacitance-to-Phase Transducer

Fig. 7 represents a more realistic model of the circuit, including the input impedance of the amplifier, the frequency dependent open-loop gain $A(f)$ of the amplifier and the inter-winding capacitance $C_w$ of the inductor.

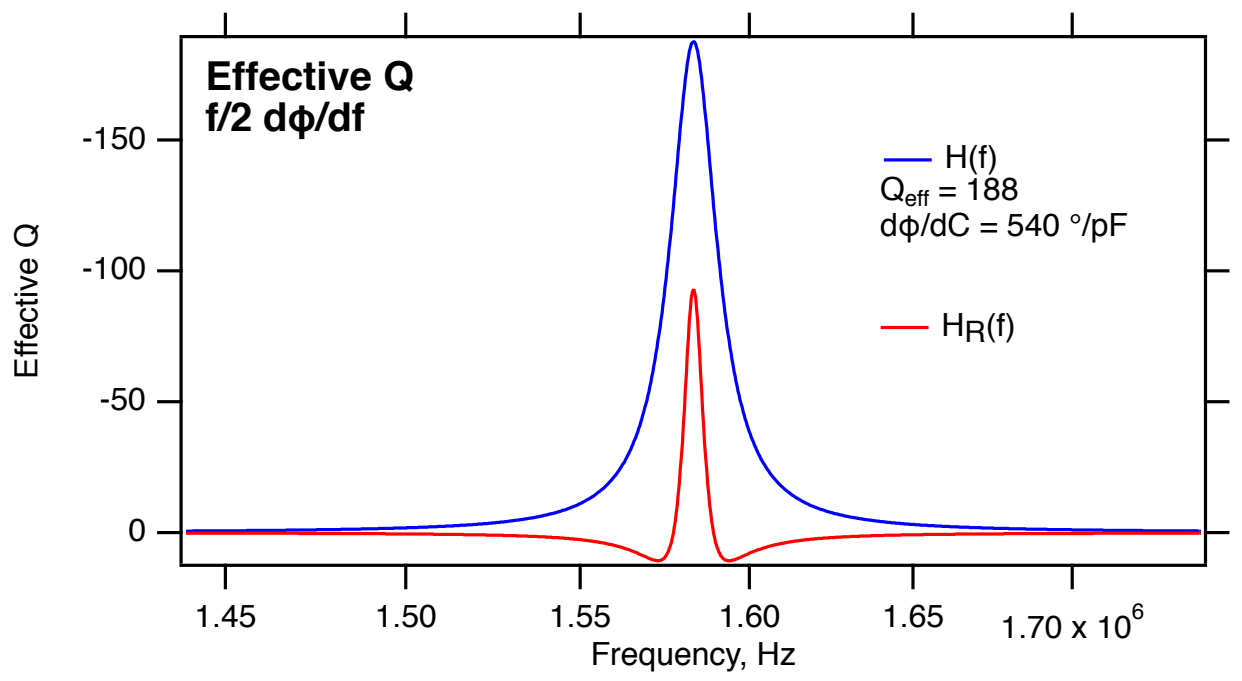

**Figure 6.** Effective Q-factor of the notch anti-resonance (red) and the corresponding APF (blue)

The differential mode input impedance of the amplifier consists of a resistor $R_d$ in parallel with a capacitor $C_d$. The common mode impedance is similar and is quoted in the data sheets for the two inputs connected in parallel, and both the common mode capacitance $C_c$ and resistance $R_c$ are taken to be half of quoted values. The circuit can be solved using Kirchhoff's law to construct an equation for each of the nodes in the circuit. An algebraic solution is easily obtained using computer-aided algebra. However, the resulting expressions are too cumbersome to be of any value in addition to numerical evaluation. Solutions for the transfer function corresponding to each of the sources were readily obtained and plotted.

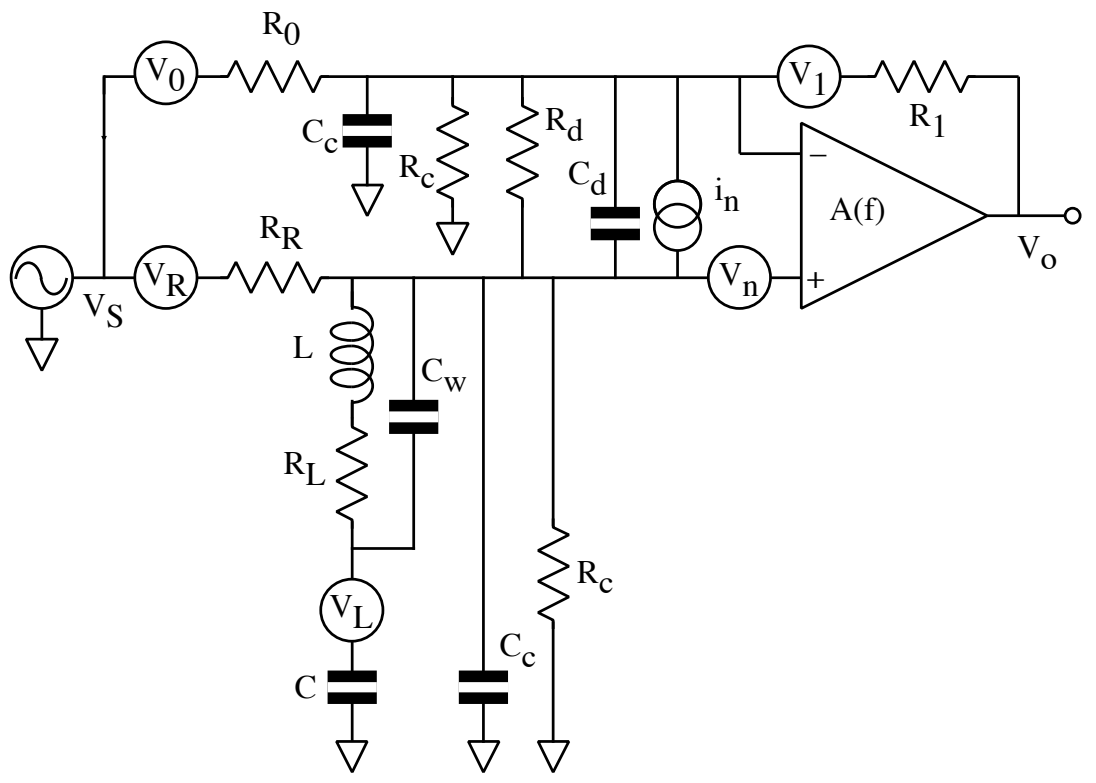

**Figure 7**. The extended model of the transducer circuit. This is a complete linear model, including all noise sources and the "imperfections" in both the amplifier and inductor

The transducer impedance in the full model is much more complicated because of the shunting effect of both the winding capacitance and the common mode input impedance. Moreover, it is also affected by the feedback from the amplifier, acting through the differential input impedance. Ignoring the latter effect, the impedance is shown in Fig. 8. This is a highly zoomed-in view, and Fig.

9 shows the same impedance over a wider frequency range, to encompass the resonance peak introduced by the shunt capacitance.

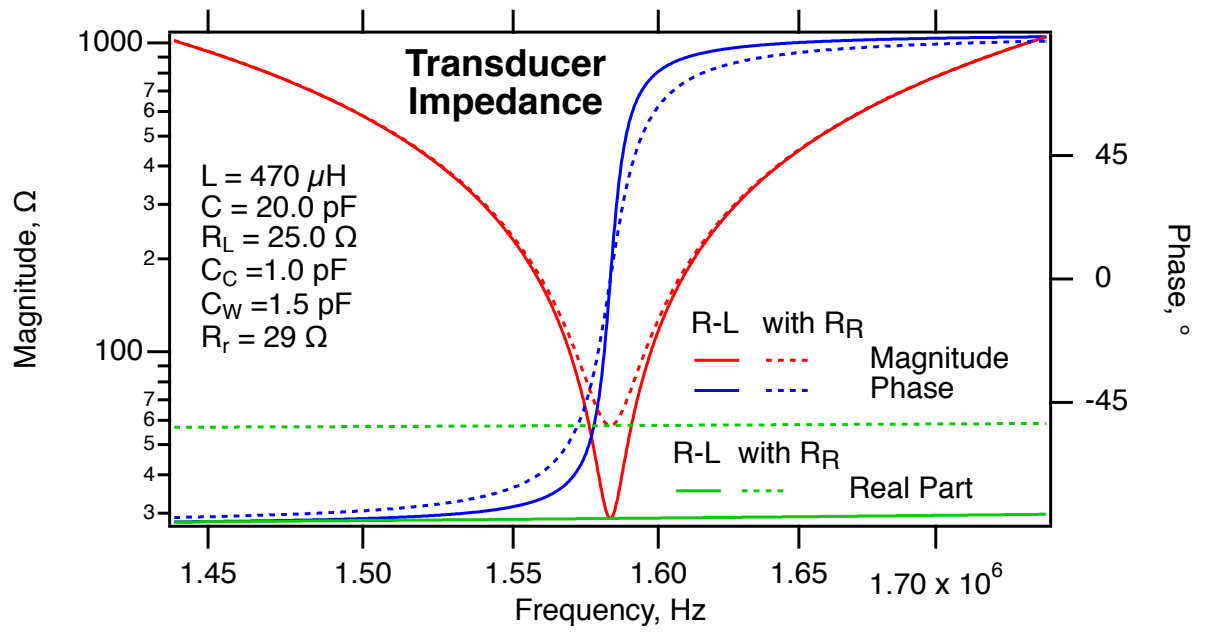

**Figure 8**. The transducer impedance modified by the effect of the winding and input capacitances

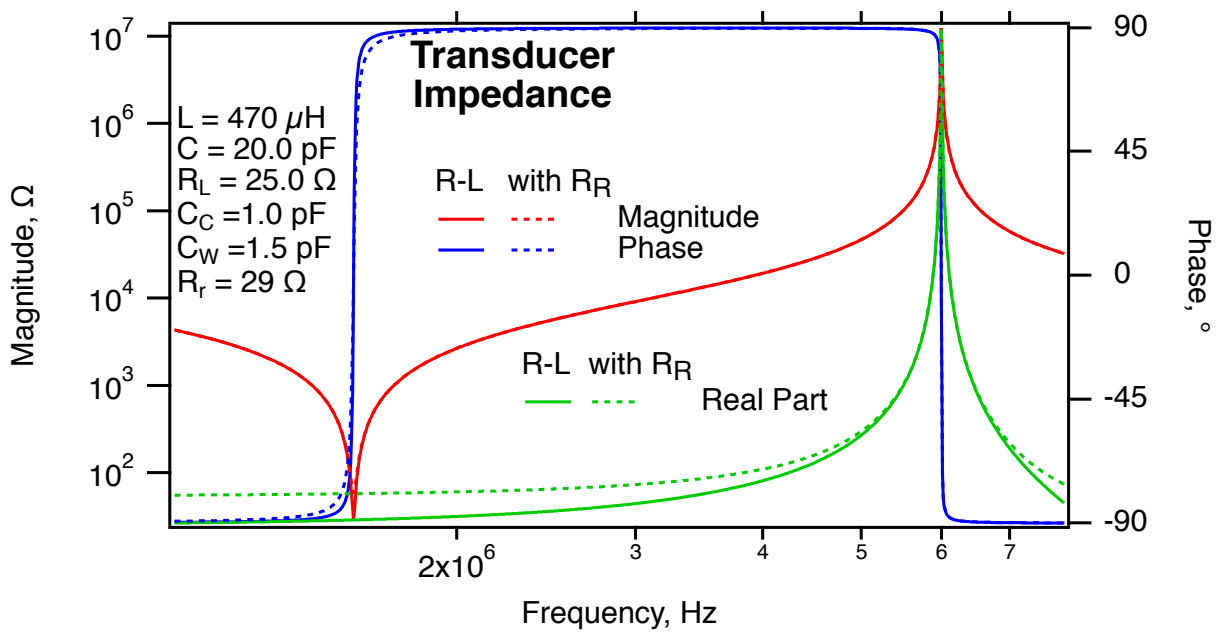

**Figure 9**. A zoomed out view of the impedance in Figure 8

This peak is removed by less than a decade in frequency from the minimum, and this raises the impedance at the notch. From Fig. 8, it is evident that the real part of the impedance is increased from 25 Ω to 28.9 Ω and the value of $R_R$ must be increased accordingly to maintain the optimum 50% depth of the notch in $H_R$. This is shown in Fig. 10 and appears to be very similar to the ideal model shown in Fig. 3.

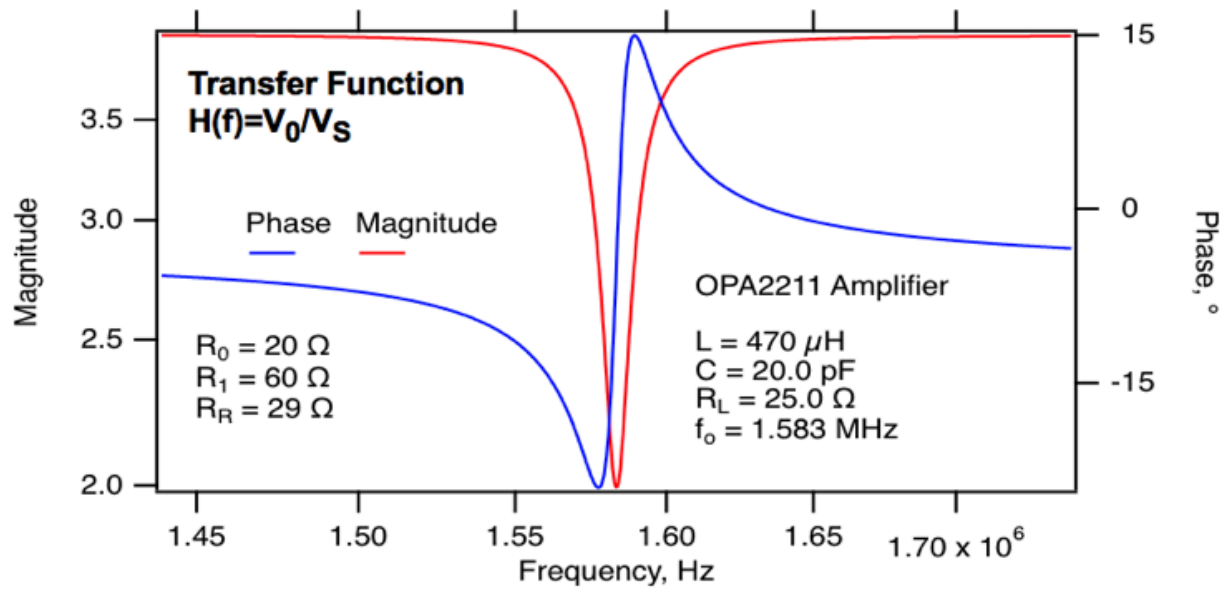

**Figure 10**. The transfer function for the extended circuit model

However, the transfer function is no longer a perfect notch, with the pole and zero frequencies slightly separated, but this is imperceptible from the graph as the separation is only 1.5 Hz. This can be ascertained from the algebraic solutions, by extracting the coefficients of both the numerator and denominator polynomials and using a numerical root solver to determine the poles and zeros. These are shown in Fig. 11, although in this view too it is impossible to resolve the 1.5 Hz frequency split. However, it can be seen by examining the numerical values of the respective roots.

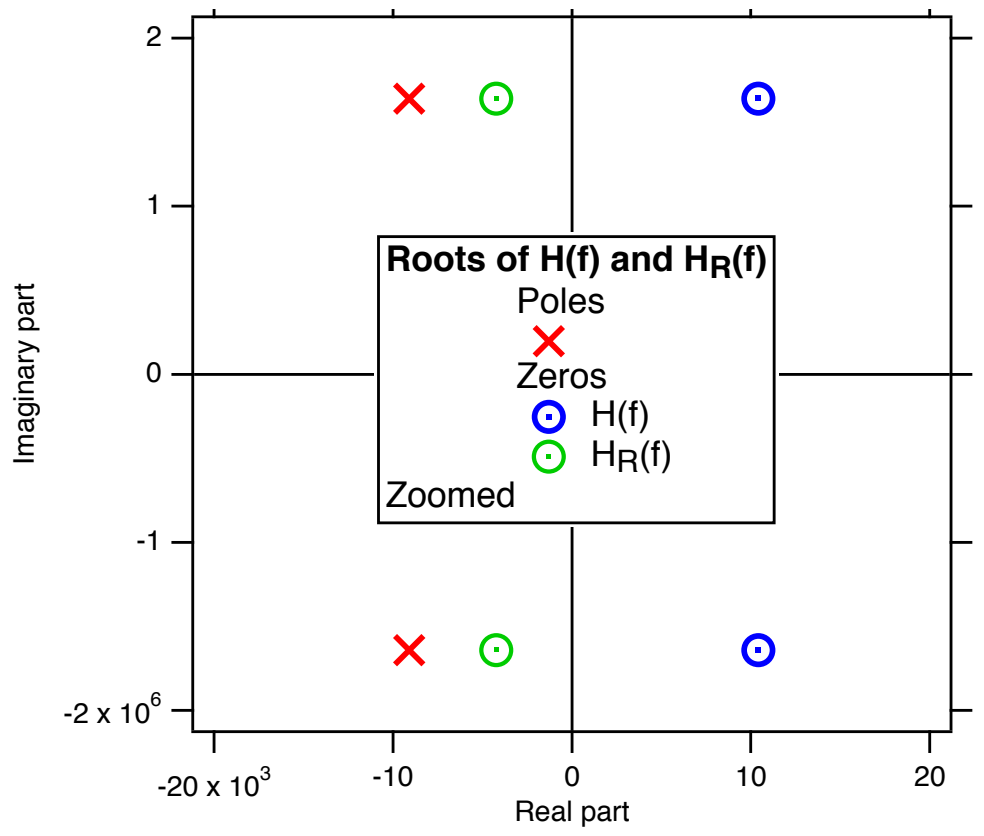

**Figure 11**. Poles of the transfer functions (red) and zeros for $H_R$ (green) and $H$ (blue)

In Fig. 11, the red crosses correspond to the complex conjugate poles (roots of the denominator polynomial) and are common to all the transfer functions as they define the characteristic equation of the system. The green circles correspond to the complex conjugate zeros (roots of the numerator polynomial) of the transfer function $H_R$. This is typical of the transfer function of a second order notch filter, where the depth of the notch is given by the ratio of the two Q factors (pole and zero), whereas the slope of the phase is proportional to the difference between them. The blue circles correspond to the zeros of the APF and, as expected, are a mirror image of the poles. Fig. 11 is a zoomed-in view because the system is fifth-order, and the other roots would compress the scale too much if included. These other roots were sufficiently far removed to have only a small effect on those displayed in Fig, 11. The corresponding APF is shown in Fig. 12. Here the gain has been iteratively tuned to flatten the transfer function as much as possible, with a resolution of approximately 0.01% for the gain value. Despite this careful tuning, small residual ripple remains in the transfer function, which is due to the frequency mismatch of the notch dip. The residual ripple is quite small, on the order of 0.1%.

## 7. Noise of the Transducer Including All Sources

The solution to the circuit equations provides the transfer function corresponding to each of the noise sources, and the respective contributions to the total noise are plotted in Fig. 13. Assuming that the noise sources are independent, their respective PSDs can be summed to obtain the total noise $S_V$, which is also shown. Of course, only the total noise can be observed experimentally and the individual contributions cannot be separated. However a knowledge of their theoretical contributions is helpful for optimizing the circuit parameters.

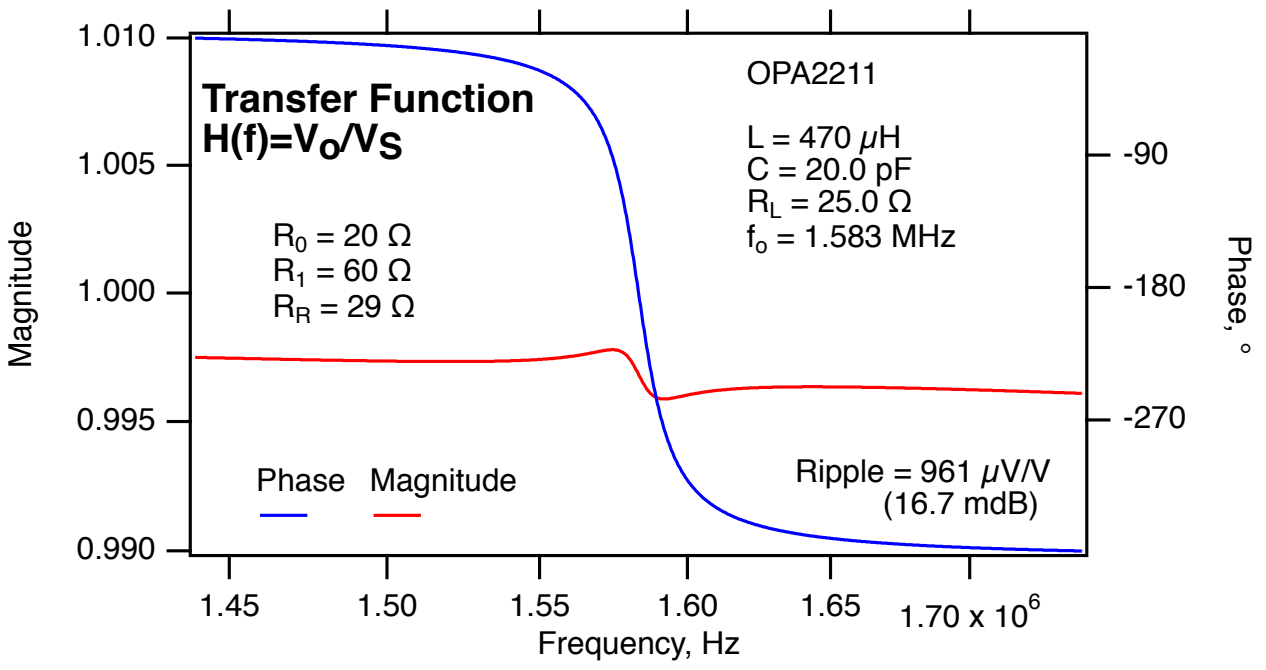

**Figure 12.** The APF tuned as well as possible but showing the residual ripple due to the slight mismatch between the frequency of the poles and the zeros

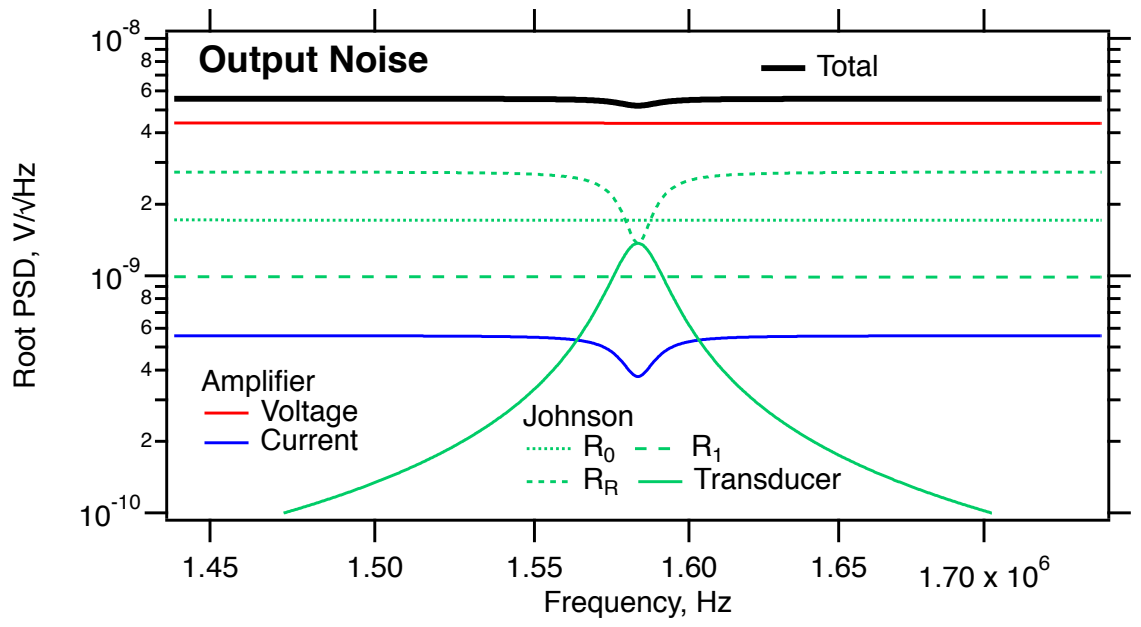

**Figure 13**. The PSD of the noise at the output of the amplifier and its constituent contributions

To convert the total output noise depicted in Fig. 13 into equivalent capacitance or displacement noise one needs a generalization of Equations (6) and (7), which apply only at the resonance frequency. The broadband versions are as follows

$$S_C = \left(\frac{C_o}{Q_{eff}HV_p}\right)^2 S_V \tag{9}$$

$$S_x = \left(\frac{x_o}{Q_{eff}HV_p}\right)^2 S_V \tag{10}$$

where $V_p$ is the amplitude of the pump signal $V_S$, $Q_{eff}$ is the function plotted in Fig. 9 and *H(f)* is the transfer function plotted in Fig. 12. The PSD of the effective displacement noise $S_x$ , when the circuit is used as a position transducer, is shown in Fig. 14. This estimate assumes that the equilibrium spacing between the electrodes of C is 100 µm, which although it is a non-trivial specification, has been readily achieved in practice, and even smaller values have been attained with careful alignment. This makes the noise estimate reasonably conservative. On the other hand, any stray capacitance in parallel with C, which is not in the model would increase the amplitude of the noise in proportion to $\frac{C+C_S}{C}$ , where $C_S$ is the undesired stray capacitance.

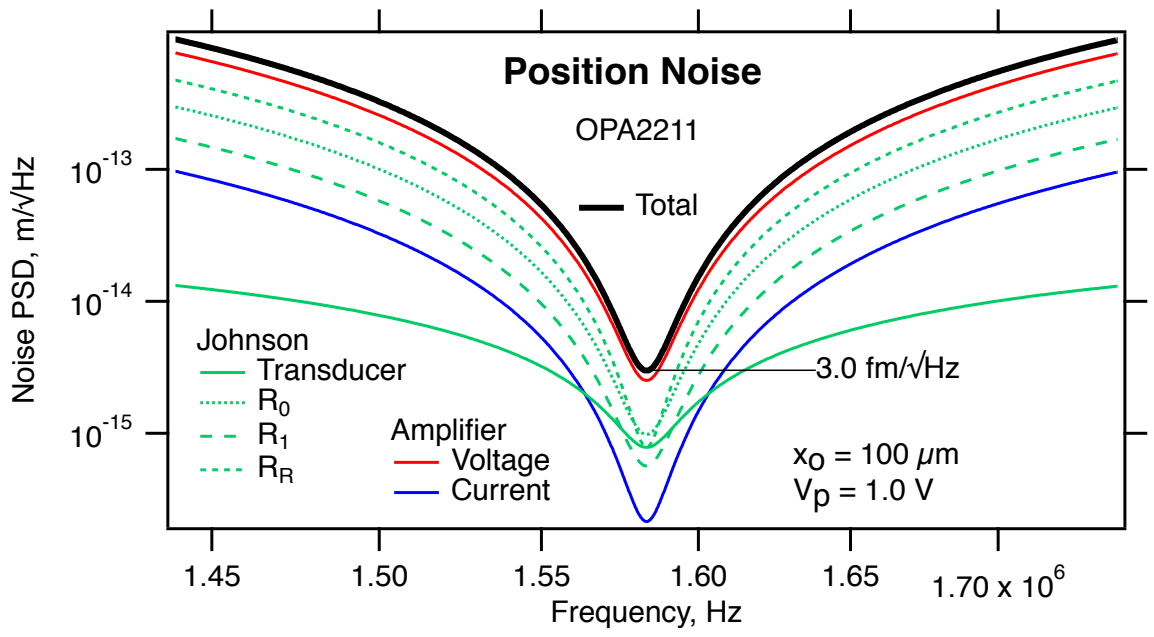

**Figure 14**. The PSD of the effective position noise corresponding to the output noise of Figure 13

Careful attention to the circuit layout is required to minimise this problem. It should be noted that exactly the same noise is achieved without connecting the signal to $R_0$ (as in Fig. 3 rather than in Fig. 7), but at the expense of the APF function. In this case, the transfer function is that of Fig. 5, with a 6dB notch rather than that of Fig. 12, with only 0.017 dB ripple.

## 8. Experimental Verification – The Test Circuit

The test set up consisted of a specifically designed test circuit board coupled to a HP 89441A 2-channel dynamic signal analyzer (DSA) as shown in Fig. 15. The primary components here are the inductor L (Q4 series ferriteless 470 uH +/- 1%, CoilQ Inc., USA), and sensing capacitor $C_0$ (Cornell Dublier model CM05CD, silver-mica, 18 pF, 500 V). Resistors $R_0$ and $R_R$ are both 50-ohm multiturn trim pots (BOURNS 3214J-1-500E 10%) and resistor $R_1$ was a fixed 150 Ω metal foil film resistor (VISHAY 0.01 %). $R_2$ and $R_3$ are standard 200 ohm 1% resistors that can provide 6 dB gain compensation, if necessary, due to 50 Ω input termination for the input buffering section of the test circuit board. The buffering part A and the sensing part B are made of a dual OPA2822 2.0 nV/√Hz unity-gain stable operational amplifier. Switches SW1 and SW2 allow for the selection of four possible test modes and are described in Section 9. Capacitor $C_1$ (0.1 pF +/- 0.01pF, KYOCERA thin film 100 V 0201) together with jumper J1 and the associated stray capacitances is used for a sensitivity estimate, and this is described in Section 9.4.

## 9. Experimental Verification – The Data

### *9.1 The transfer function* $H_R(f)$

The observed transfer function $H_R$ is shown in Fig. 17. The measurement was made using the HP89441A DSA, configured to measure frequency response and with the signal source set to provide a band-limited periodic chirp.

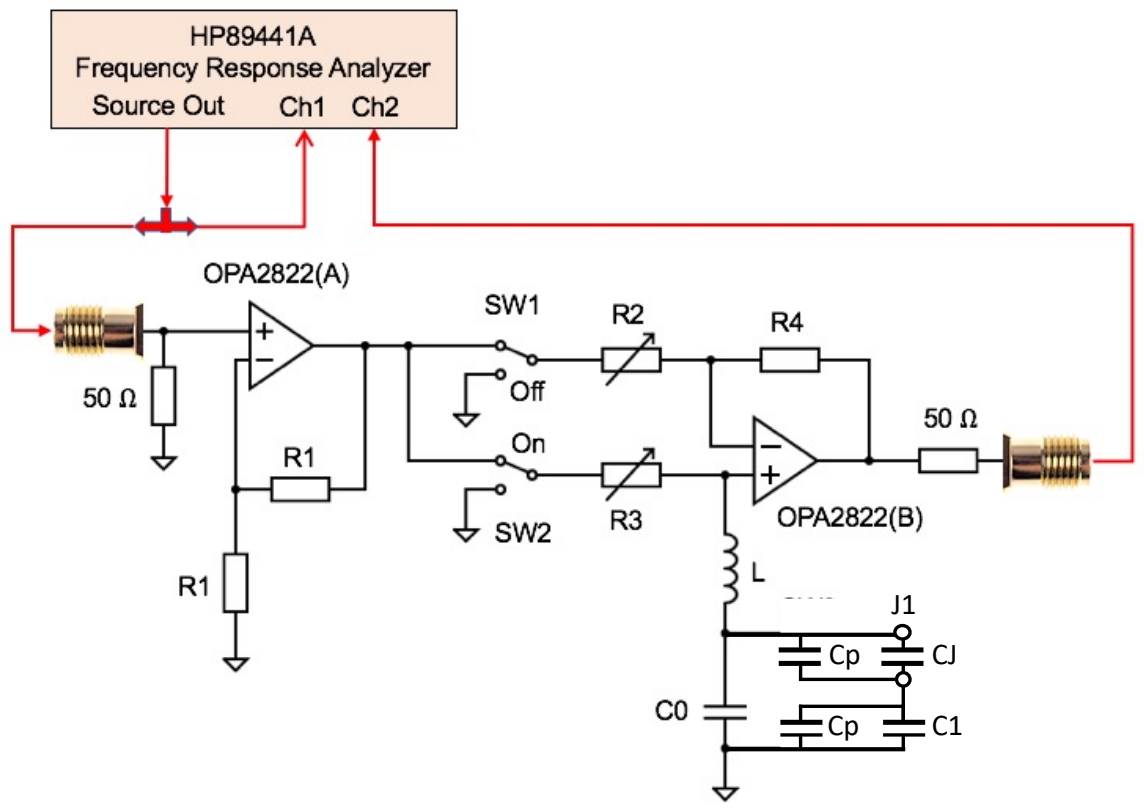

**Figure 15**. Schematic diagram of the test apparatus

A photograph of the apparatus is shown in Fig. 16 below

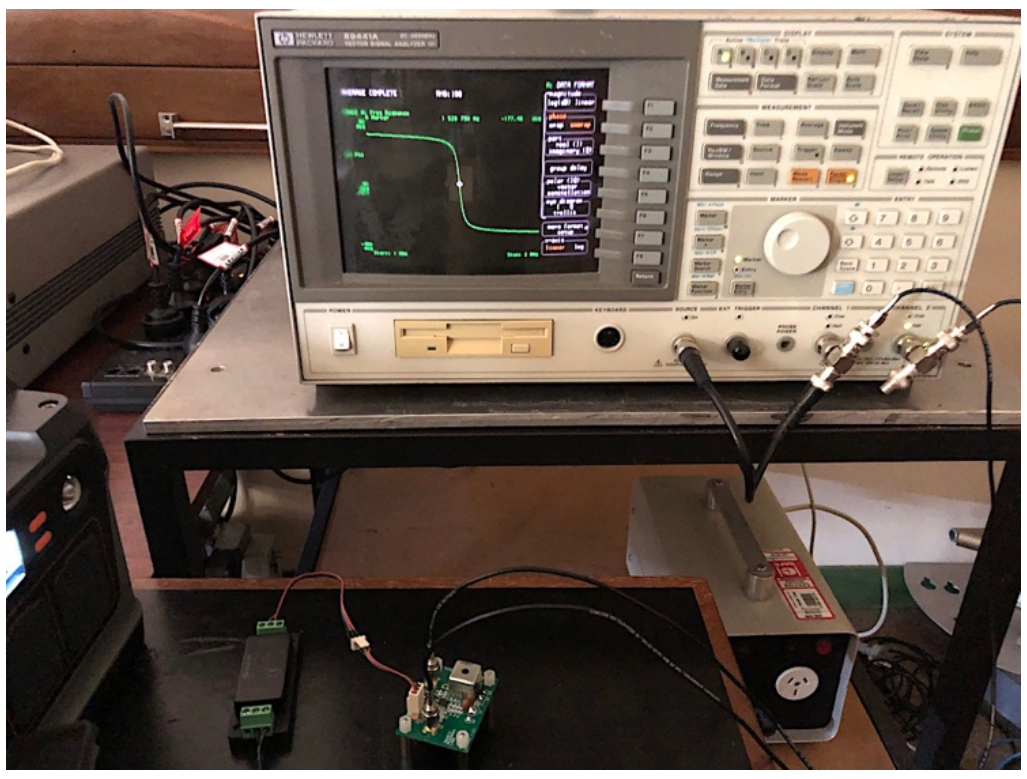

**Figure 16**. Photograph of the HP89441A DSA showing the phase vs frequency response of the test board under evaluation

This measurement is required to adjust the value of $R_R$ to produce the desired notch depth of 50% and is made with SW1 set to off and SW2 set to on. A least-squares fit to the data using the model for $H_R$, plotted in Fig. 17, shows excellent agreement between the data (plotted as dots) and the best fit (plotted as a smooth curve), and the two traces cannot be distinguished because their separation is smaller than the thickness of the line. However, this can be resolved in the plot of the residuals at the top of Fig. 17, indicating that the residual error is approximately 0.1% rms. The phase data are also plotted on the graph, together with the best fit curve, which are also in excellent agreement because the fit was performed using the complex data and fitting jointly to the real and imaginary parts. However, the residuals were only slightly worse when fitted only to the magnitude. The values of both $R_0$ and $R_R$ were known and fixed for fitting. For other parameters, whose values are accurately known, these agree with those estimated by the fit. Some, including $C$ and $R_L$ and the input capacitance of the amplifier are difficult to measure independently, and their values are determined by the curve fitting. The resulting input capacitance is slightly larger than that quoted for the amplifier, and this is attributed to the stray capacitance on the circuit board.

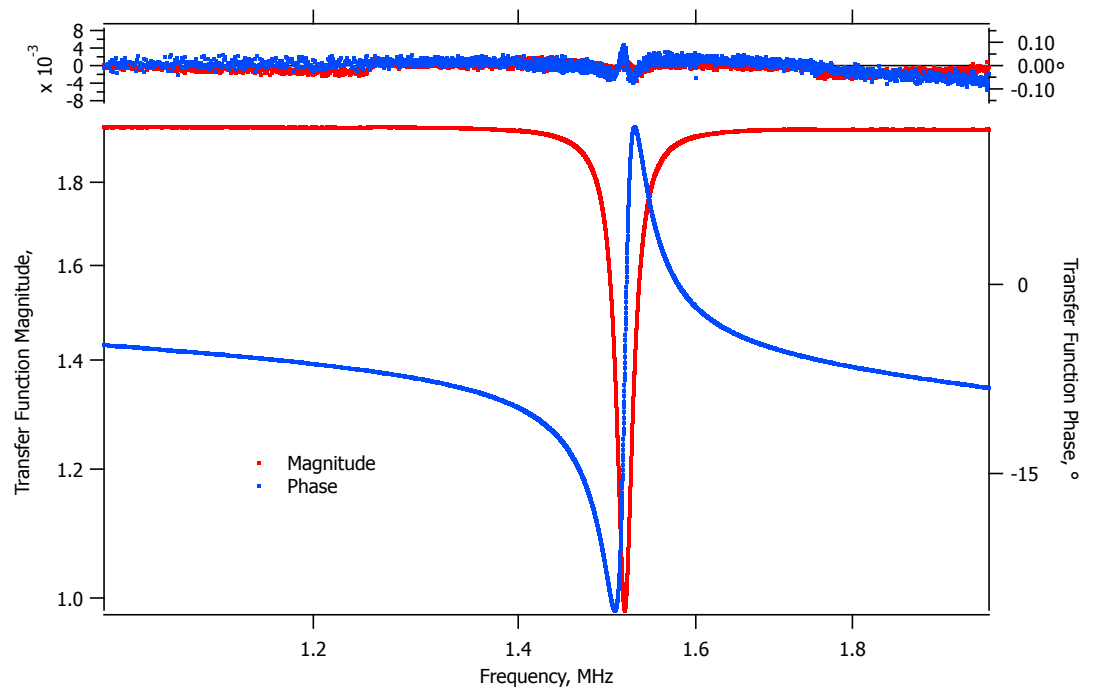

**Figure 17**. An example of the measured transfer function $H_R$ (the diagnostic configuration), but with the circuit parameters not optimally tuned, showing the least-squares fit

### *9.2 The transfer function of the APF*

The observed transfer function corresponding to the APF operating mode, with both SW1 and SW2 set to on is shown in Fig. 18, together with the result of performing the least-squares fit to the corresponding model and again showing excellent agreement.

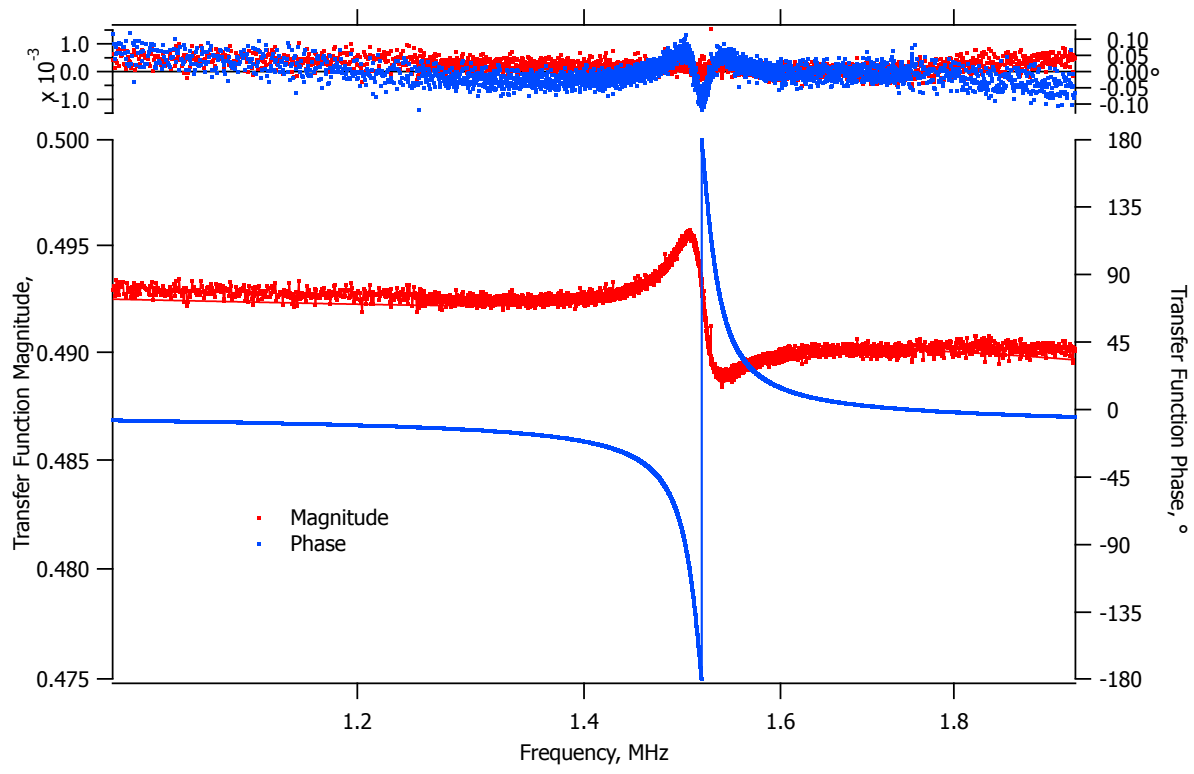

**Figure 18**. An example of the measured transfer function for the APF configuration but with the circuit parameters not optimally tuned, showing the least-squares fit (phase wrapped mode)

Here, the vertical scale of the magnitude plot is greatly magnified. Thus, it is possible to see the difference between the data and the fit, but the fit is still excellent with residuals smaller than 0.1% rms. The APF was not perfectly flat, and the ripple was consistent with the unavoidable stray capacitance described in the previous section. Further tests need be conducted, using circuit boards with an improved layout to reduce the stray capacitance and also using inductors with even larger Q values.

### *9.3 The Noise of the Test Circuit*

The noise of the circuit was measured with both SW1 and SW2 set to off and as shown in Fig. 19. Because the gain of the amplifier is necessarily modest (at approximately 4), the noise at the output is not significantly greater than the input noise of the DSA and many averages are required to resolve the circuit noise. In addition, it is necessary to subtract the noise of the DSA, which must also be measured with many averages. The processed noise is shown in Fig. 19 together with the noise of the DSA for comparison.

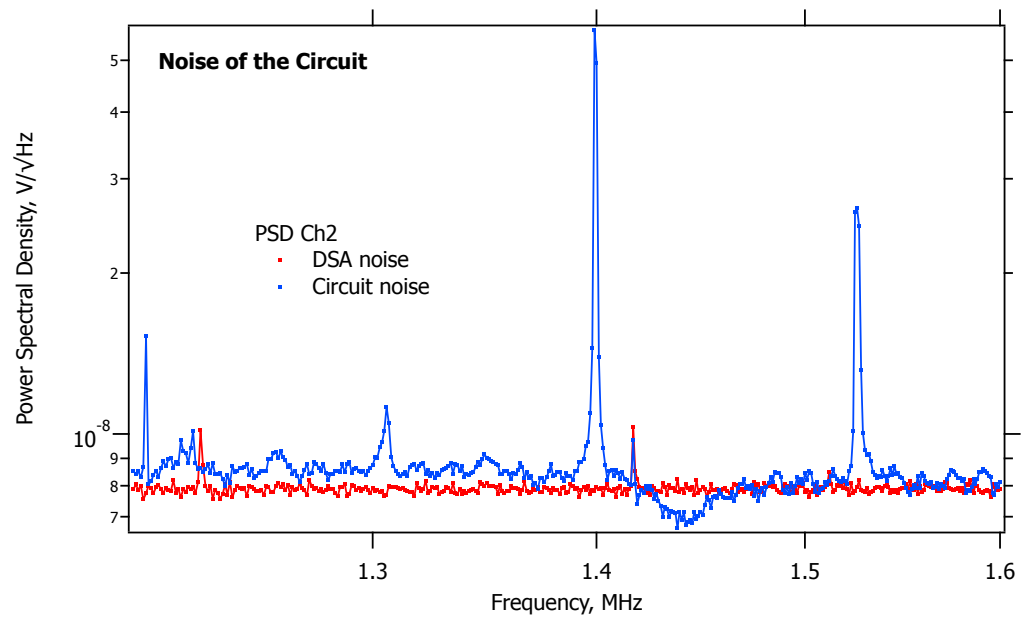

**Figure 19**. The measured output noise of the test circuit

Ignoring the laboratory-based interference peaks, the noise follows the model shown in Fig. 13, with the dip at the resonance frequency. The interference peaks were very narrow and sufficiently far from the resonance of the circuit to be ignored.

### *9.4 Phase Shift Versus Small Capacitance Shift Measurement*

A simple method of estimating the sensitivity $\frac{d\varphi}{dC}$ by measuring the phase shift of the pump signal caused by adding capacitor $C_1$ (0.1 pF ± 0.01pF) to the sensing capacitor $C_0$ (implemented for the tests as a fixed silver mica capacitor of measured value 18.3 pF) is shown in Fig. 15 above. This was done by connecting $C_1$ in parallel to $C_0$ by installing jumper *J1*, which was initially removed. However, the resulting effective capacitance is not simply the sum of $C_0+C_1$ connected in parallel. The terminals of the jumper have a stray capacitance $CJ$ that was estimated numerically and accurately measured using a commercial LCR meter (SANWA LCR700, made in Japan) and found to be 0.12±0.01 pF, and the stray capacitance between the tracks of the PCB was estimated to be $Cp$ ≈0.2 pF [28], [29]. The resulting effective change in capacitance is calculated by using Equation (10)

$$\Delta C = \frac{(C1+Cp)^2}{C1+2Cp+CJ} \approx 0.15\, pF \qquad (10)$$

The phase shift caused by installing the jumper J1 is shown in Fig. 20 and the frequency response of the circuit was measured for the jumper J1 either removed or installed. Numerical differentiation is used to obtain the slope $\frac{d\varphi}{df}$ of the phase as a function of frequency and hence the corresponding slope $\frac{d\varphi}{dC}$ of the phase as a function of capacitance as in Equation (4), where $\frac{dC}{df} \approx \frac{2C}{f_0}$. One has

$$\frac{d\varphi}{dC} = \frac{d\varphi}{df}\left(\frac{dC}{df}\right)^{-1} = \frac{d\varphi}{df}\,\frac{f_0}{2C} \qquad (11)$$

The measured phase shift corresponds to ≅ 5.1 rad/pF and is in good agreement with the value of ≅ 5.0 rad/pF computed from the frequency response data. This agreement is as close as possible when using such small capacitances, which are very difficult to measure precisely.

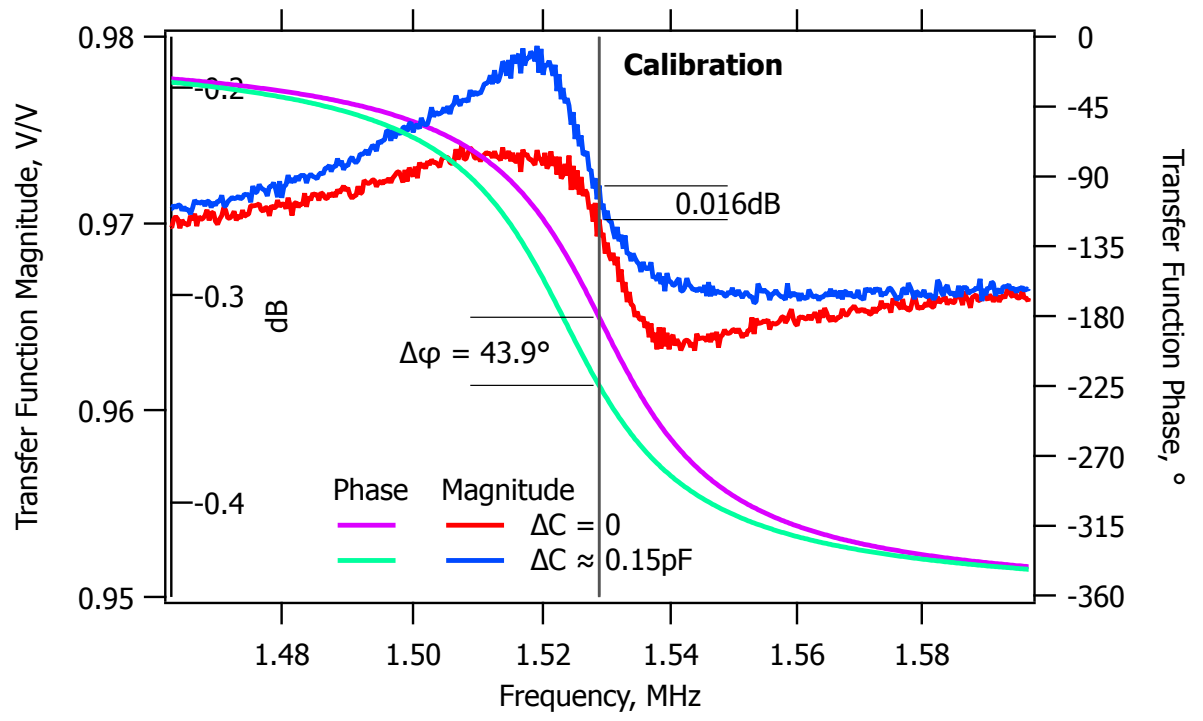

**Figure 20**. Measurements showing the phase change of ~ 293 deg/pF ( ~ 5 rad/pF ) produced by adding ~0.15 pF to the sensing capacitor. The corresponding amplitude shift is measured to be approximately 0.1 dB/pF and maximum ripple ~ 0.01 dB/kHz in the frequency domain

It is important to emphasise that the proposed capacitance-to-phase transducer does not represent a complete capacitive sensor design. The latter depends on multiple options for the back-end constituents that convert the phase modulated carrier signal into voltage modulated output signal which can then be digitized, recorded and visualised by proper means ( see the recent excellent review [24] and cited references therein ). A few possible realisations of the complete capacitive sensor designs based on this particular capacitance-to-phase transducer will be discussed in a different publication elsewhere. It is also important to note that the optimum performance of the proposed capacitance-to-phase transducer is located approximately within 500 kHz – 10 MHz carrier frequency range and cannot be compared with other types of transducers operating at much lower or much higher carrier' frequencies such as microwave or optical transducers. The proposed capacitance-to-phase transducer has been specifically developed for use as a front-end capacitive read-out in advanced miniaturised gravity gradiometers [30]. It demonstrated outstanding performance during a recent out-of-the-lab trial in a harsh outback environment in Western Australia. The results of the latter will be reported elsewhere.

## Summary

For the first time, the classical series resonant circuit, combined with a modified All-Pass filter architecture, forms the basis for a sensitive, cost-effective, single-chip capacitance-to-phase transducer. It can be naturally incorporated into well-known carrier phase noise cancelling signal processing schemes, such as interferometers, flip-flops and lock-in amplifiers. Containing only a few critical precision electronic components, it promises unparalleled sensitivity for operation at room temperature and potentially in elevated-temperature environments. If configured as a displacement-to-phase converter using moving plate sensing capacitors, the predicted position noise, if properly tuned, is about 3 fm/√Hz which is similar to or lower than has been demonstrated in the best room temperature capacitive resonant bridges reported to date ( see [9] ). The circuit is specifically designed for grounded sensing capacitors which is a significant advantage when used in mechanical displacement sensors. The transducer is configured such that the transfer function is a close approximation to an All-Pass filter eliminating the effects of AM modulation and simultaneously providing a much higher capacitance-to-phase conversion rate. The presented theoretical analysis and modelling of the proposed capacitive sensing metrology closely matched the experimental results obtained with a simple test circuit and those skilled in the art can make one for independent assessment. The circuit and its possible modifications have been protected by US Patent Application 18230348.

## Acknowledgements

Author van Kann wishes to thank John Winterflood of OzGrav Group at the UWA for many useful, illuminating, and soul-searching discussions. Author Veryaskin would like to thank Marc Kaye of CadLink (Western Australia) for many years of his professional service to the Trinity Research Lab by developing PCBs and assembling precision electronic units.

This work was supported in part by Lockheed Martin Corporation, RMS Precision Navigation – Gravity Systems, Niagara Falls, NY, 65M1

## Authors Contribution

Although both authors collaborated on every aspect of this work, Veryaskin invented the original concept of basing the capacitive transducer on a variant of the resonant circuit. He designed and constructed the apparatus and performed the experiments. Van Kann performed a detailed theoretical analysis of the circuit which resulted in additional modification of the original concept, produced the simulated transfer functions, processed the experimental data, and constructed the full model of the transducer noise.

## References

[1] Puers R 1993 Capacitive sensors: when and how to use them Sensors and Actuators A 37(38) 93-105

[2] Xiaohui Hu and Wuqiang Yang 2010 Planar capacitive sensors – designs and applications Sensor Review 20(1) 24–39

[3] D'Allesandro A, Scudero S and Vitale G A 2019 Review of the Capacitive MEMS for Seismology Sensors 19 3093

[4] Penfield Jr and Rafuse R P 1962 Varactor Applications The Massachusetts Institute of Technology Press Cambridge (USA)

[5] Noras M A 2014 Electric field detection using solid state variable capacitance Proc. ESA Annual Meeting on Electrostatics 12 p.

[6] Neubert H K P 1975 Instrument Transducers Oxford University Press Oxford (UK)

[7] Oide K, Hirakawa H and Fujimoto M K 1979 Search for gravitational radiation from the Crab pulsar Phys. Rev. D 20 2480-2483

[8] Mendousse J S, Goodman P D and Cady W G 1950 A Capacitance Bridge for High Frequencies Rev. Sci. Instrum. 21 1002-1009

[9] Hu M, Bai Y Z, Zhou Z B, Li Z X and J Luo 2014 Resonant frequency detection and adjustment method for a capacitive transducer with differential transformer bridge Rev, Sci, Instrum. 85, 055001

[10] Dan Bee Kim, Hyung Kew Lee and Wan-Seop Kim 2017 An impedance bridge measuring the capacitance ratio in the high frequency range up to 1 MHz Meas. Sci. Technol 28 025014

[11] Josselin V, Touboul P and Kielbasa R 1999 Capacitive detection scheme for space accelerometers applications Sensors and Actuators 78 92–98

[12] De Bra D B 1976 Control Requirements of Space Relativity Experiments IFAC Proceedings 9 142-160

[13] Jacobs E D 1968 New developments in servo accelerometers Inst. Env. Sci. 14th Ann. Tech. Meet. 1-7

[14] Braginsky V B and Manukin A B 1977 Measurement of weak forces in physics experiments Edited by David H. Douglass University of Chicago Press

[15] Blair D G, Ivanov E N, Tobar M E, Turner P J, van Kann F and Heng I S 1995 High Sensitivity Gravitational Wave Antenna with Parametric Transducer Readout Phys. Rev. Let. 74 1908-1911

[16] Nguyen C and S. Kim S 2012 Analysis of RF Interferometer In: Theory, Analysis and Design of RF Interferometric Sensors Springer Briefs in Physics Springer (NY)

[17] Ivanov E N, Tobar M E and Woode R A 1998 Microwave interferometry: application to precision measurements and noise reduction techniques IEEE Transactions on Ultrasonics, Ferroelectrics and Frequency Control 45(6) 1526-1537

[18] Rubiola R and Giordano V 2002 Advanced interferometric phase and amplitude noise measurements Rev. Sci. Instrum. 73(6) 2445-2457

[19] Ashkan Ashrafi and Hossein Golnabie 1999 A High Precision Method for Measuring very Small Capacitance Changes Rev. Sci. Instrum. 70(8) 3483–3487

[20] Matko V and Milanović M 2014 Temperature-compensated capacitance–frequency converter with high resolution Sensors and Actuators 220 262–269

[21] Inigo RM 1969 A Single Operational Amplifier Notch Filter IEEE Proceedings (Letters) 57 727

[22] Upadhyay A and Pal K 2015 Grounded capacitor bandpass filter using operational amplifier pole with low component sensitivity International Conference on Computing, Communication & Automation Greater Noida India 1417-1420

[23] Zumbahlen H Allpass Filters Analog Devices Mini Tutorial MT-202

[24] Kanoun Olfa, Ahmed Yahia Kallel , Ahmed Fendri 2022 Measurement (IMEKO) Measurement Methods for Capacitances in the Range of 1 pF–1 nF: A Review 195 111067

[25] Johler R J and Walters L C Mean Absolute Value and Standard Deviation of the Phase of a constant Vector Plus a Rayleigh-Distributed Vector 1959 Journal of Research of the National Bureau of Standards 62(5) 183-186

[26] van Kann F J (to be published)

[27] Delyiannis T, 1968 High-Q Factor Circuit with Reduced Sensitivity Electronic Letters 4 577

[28] Jackson J D 1975 Classical Electrodynamics (2nd ed.) New York John Wiley & Sons

[29] Tool Box: https://www.emisoftware.com/calculator/coplanar-capacitance/

[30] Veryaskin A V 2021 Gravity, Magnetic and Electromagnetic Gradiometry: Strategic Technologies in the 21st Century (2nd Edition) IOP Publishing Ltd. 190 pp https://iopscience.iop.org/book/mono/978-0-7503-3803-5